\documentclass[a4paper]{jpconf}
\usepackage{amsmath}
\usepackage{graphicx}
\usepackage{amssymb}
\usepackage{bbm}
\usepackage{graphicx}

\newcommand{\cC}{\mathcal{C}}
\newcommand{\cK}{\mathcal{K}}
\newcommand{\cL}{\mathcal{L}}
\newcommand{\cM}{\mathcal{M}}
\newcommand{\cN}{\mathcal{N}}

\newcommand{\Zom}{\mathbbm{Z}}

\newcommand{\SL}{\mathrm{SL}}

\newcommand{\I}{i}

\newcommand{\vb}{\bar{v}}
\newcommand{\wb}{\bar{w}}

\newcommand{\vr}[1]{{\vec{r}\,}^{#1}}

\def\be{\begin{eqnarray}}
\def\ee{\end{eqnarray}}

\begin{document}
\begin{flushright} \small
ITP--UU--07/57 \\ SPIN--07/43 \\SPHT--T07/136
\end{flushright}

\title{Recent results in \\ four-dimensional non-perturbative string theory
\footnote{Based on talks given at the European Physical Society HEP 2007 Conference, 19--25 July 2007, Manchester, 
England and at the Fifth Simons Workshop in Mathematics and Physics, 30 July -- 31 August 2007, Stony Brook, USA.}
}

\author{Frank Saueressig $^{1,2}$}

\address{$^1$ Service du Physiqu\'e Theoriqu\'e, Orme des Merisiers, CEA/Saclay, \\ 91191 Gif-sur-Yvette Cedex, France}

\address{$^2$ Institute for Theoretical Physics and Spinoza Institute, Utrecht University, \\ 3508 TD Utrecht, The Netherlands }

\ead{Frank.Saueressig@cea.fr}

\begin{abstract}
We review recent progress in understanding non-perturbative instanton corrections to the hypermultiplet moduli space in  type II string compactifications on Calabi-Yau threefolds. 
\end{abstract}
%
\section{Introduction}
\begin{figure}[t!]
\setlength{\unitlength}{1cm}
\begin{center}
\begin{picture}(16,10)
\thicklines \put(3.5,9.4){\makebox(2,0.6)[c]{Hypermultiplet sector
$\cM_{\rm HM}$}} \put(12.5,9.4){\makebox(2,0.6)[c]{Vector
multiplet sector $\cM_{\rm VM}$}}
\put(0.5,8.2){\makebox(2,0.6)[c]{IIA / $X$}}
\put(7,8.2){\makebox(2,0.6)[c]{IIB / $Y$}}
\put(10.75,8.2){\makebox(2,0.6)[c]{IIA / $Y$}}
\put(14.25,8.2){\makebox(2,0.6)[c]{IIB / $X$}}
\put(10.75,0.2){\makebox(2,0.6)[c]{\{$\alpha^\prime$\}}}
\put(14.25,0.2){\makebox(2,0.6)[c]{\{$-$\}}}
\put(14.4,0.5){\vector(-1,0){1.7}}
\put(13.125,0.5){\makebox(1,0.6)[c]{mirror}}
\put(11.75,1.3){\line(0,-1){0.4}}
\put(11.75,1.3){\vector(-1,0){3.8}}
\put(9.5,1.3){\makebox(2,0.6)[c]{c-map}}
\put(5.4,1){\makebox(2,0.6)[l]{\{$\alpha^\prime$, $1\ell$\}}}
\put(6,1.7){\vector(0,1){1}}
\put(6,1.8){\makebox(2,0.6)[c]{SL(2,$\Zom$)}}
\put(0.13,2.8){\makebox(2,0.6)[l]{\{$1\ell$, A-D2\}}}
\put(5.4,2.8){\makebox(2,0.6)[l]{\{$\alpha^\prime$, D($-1$),
D1\}}}
 \put(0.75,3.5){\vector(0,1){1.1}}
\put(1,3.7){\makebox(2,0.6)[l]{e/m duality}}
\put(0,4.8){\makebox(2,0.6)[l]{ \{$1\ell$, A-D2, B-D2\} }}
\put(5.4,4.8){\makebox(2,0.6)[l]{\{$\alpha^\prime$, D($-1$), D1,
D3, D5\} }}
 \put(6,5.5){\vector(0,1){1.1}}
\put(6,5.7){\makebox(2,0.6)[c]{SL(2,$\Zom$)}}
\put(5.4,6.8){\makebox(2,0.6)[l]{\{$\alpha^\prime$, D($-1$), D1,
D3, D5, NS5\} }} \put(0,6.8){\makebox(2,0.6)[l]{ \{$1\ell$, D2,
NS5\} }}
\put(5.1,3.1){\vector(-1,0){1.7}}
\put(3.75,3.1){\makebox(1,0.6)[c]{mirror}}
\put(3.4,5.1){\vector(1,0){1.7}}
\put(3.75,5.1){\makebox(1,0.6)[c]{mirror}}
\put(5.1,7.1){\vector(-1,0){1.7}}
\put(3.75,7.1){\makebox(1,0.6)[c]{mirror}}
\put(0,8){\line(1,0){16}} \put(10.7,0){\line(0,1){1}}
\put(10.7,2){\line(0,1){8}}
\end{picture}
\end{center}
\parbox[c]{\textwidth}{\caption{\label{eins}{\footnotesize
Prospective duality chain for determining the quantum corrected LEEA of type II
strings compactified on a generic CY $X$ and its mirror partner $Y$. 
In the vector multiplet
sector there are $\alpha'$ corrections that appear on the IIA side
only and can be obtained via mirror symmetry; they comprise  worldsheet loop
and instanton corrections. The c-map transfers these
into the IIB hypermultiplet sector. In addition, there is a
one-loop $g_s$ correction, in the figure denoted by $1\ell$, determined
in \cite{RSV}. 
Imposing $\SL(2,\Zom)$ invariance
produces the non-perturbative corrections arising from D1-brane and
more general $(p,q)$-string instantons as well as D($-1$)-instantons
 \cite{Robles-Llana:2006is}. The latter naturally combine with the perturbative
$\alpha'$ and $g_s$ corrections. As shown in \cite{membrane}, applying
mirror symmetry to these corrections gives rise to the $A$-cycle
D2-brane instanton contributions on the IIA side. One might now
continue to employ various dualities that should in principle
produce the remaining quantum corrections: using electromagnetic
(e/m) duality to impose symplectic invariance will give the $B$-cycle
D2-brane instantons. Mirror symmetry will map these to the as of
yet unknown D3- and D5-brane instanton corrections on the IIB
side. Another application of $\SL(2,\Zom)$ duality then will give
rise to NS5-brane and D5--NS5 bound state instantons.
Finally, applying mirror symmetry one last time will produce the
NS5-brane corrections on the IIA side. (From \cite{membrane}.)}}}
\end{figure}
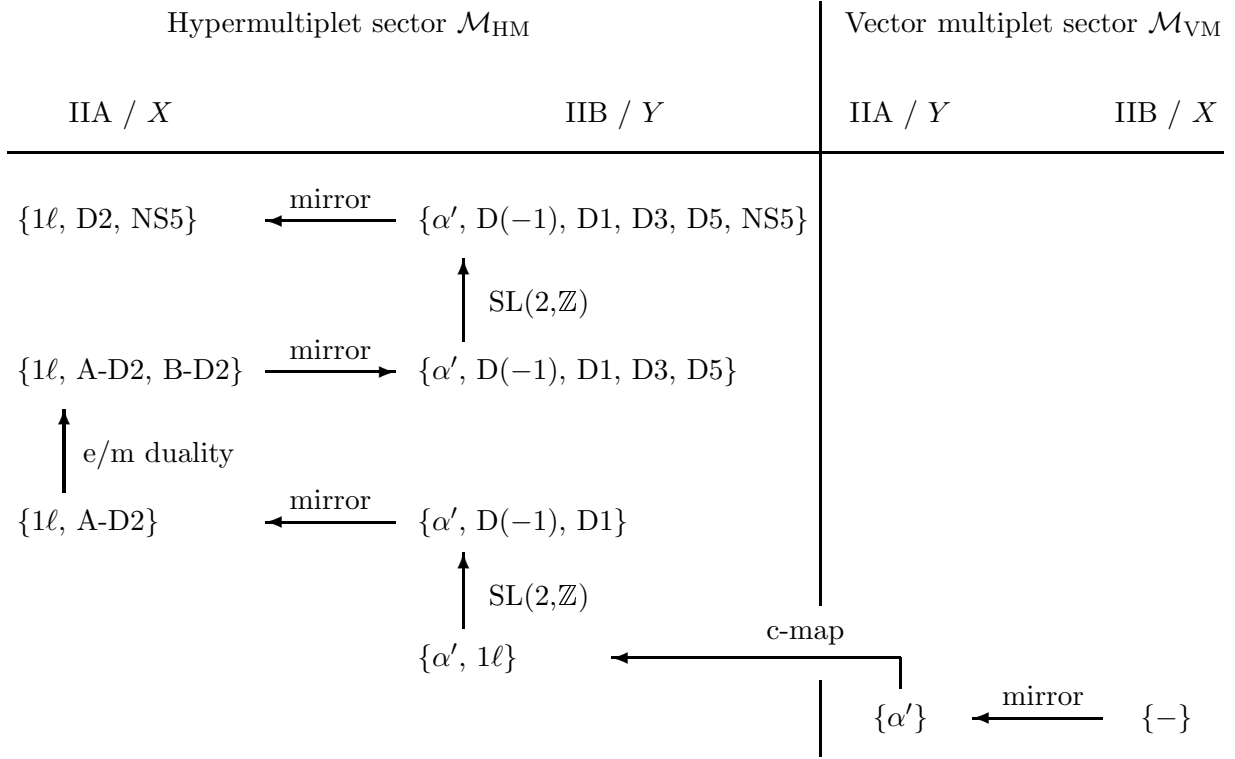

%
\setcounter{footnote}{0}
In recent years it has become clear that non-perturbative effects are an important ingredient for building semi-realistic models connecting string theory to four-dimensional low energy physics.  A setup where one might hope to understand such effects in detail is type II string theory compactified on a Calabi-Yau threefold (CY). In this case the low energy physics is captured by a $\cN = 2$, $d=4$ supergravity action and it is conceivable that the constraints coming from string dualities and supersymmetry suffice to determine the perturbative and non-perturbative corrections to the low energy effective action (LEEA), cf. Figure \ref{eins}. 

The classical LEEA can be obtained by carrying out a Kaluza-Klein reduction of ten-dimensional type II supergravity on a CY. 
The resulting massless modes organize themselves into a supergravity multiplet, $n_V$ vector multiplets (VM) and $n_H$ hypermultiplets (HM). $\cN = 2$ supersymmetry implies that the total moduli space $\cM$ of these theories factorizes into a local
product ${\mathcal M} = {\mathcal M}_{\rm VM} \otimes
{\mathcal M}_{\rm HM}$ where ${\mathcal M}_{\rm VM}$ and ${\mathcal M}_{\rm HM}$ are
parameterized by the scalars of the VM and HM, respectively. 
Furthermore, ${\mathcal M}_{\rm VM}$ must be special K\"ahler, i.e., there is a holomorphic prepotential $F$ from which the K\"ahler potential
$\mathcal{K}$ on $\cM_{\rm VM}$ can be derived. The hypermultiplet sector ${\mathcal M}_{\rm HM}$ has to be quaternion-K\"ahler 
which generically implies that the metric on $\cM_{\rm HM}$ does not admit a simple description in terms of a 
K\"ahler potential.

Beyond the classical picture the LEEA generically receives quantum corrections from the world sheet conformal field theory ($\alpha^\prime$-corrections) and in the string coupling constant ($g_s$-corrections). The $\alpha^\prime$-corrections encompass a perturbative term and world-sheet instanton corrections from (Euclidean) fundamental strings wrapping holomorphic two-cycles of the CY. The perturbative $g_s$ corrections arise from higher genus world sheets and have recently been understood in \cite{RSV}. Moreover, there are also non-perturbative $g_s$-corrections coming from Euclidean branes wrapping supersymmetric cycles of the CY \cite{BBS}. In case of type IIA (even branes) these arise from Euclidean D2-branes (membranes) wrapping special Lagrangian three-cycles of the CY or NS5-branes wrapping the CY itself. For type IIB (odd branes) one has D(-1)-, D1-, and D3-branes wrapping points, two-, and four-cycles, respectively, together with D5- and NS5-branes wrapping the CY.
With respect to these quantum corrections, the factorization of $\cM$ has the profound consequence that only
those subsectors which contain the dilaton (volume modulus)
receive $g_s$ ($\alpha^\prime$) corrections. For both IIA and IIB compactifications the dilaton sits in a hypermultiplet.
Thus determining the $g_s$ corrections to the LEEA requires understanding $\cM_{\rm HM}$, on which we will focus in the following.

\section{Describing the hypermultiplet sector by tensor multiplets}
\label{two}

When studying $g_s$ corrections, it is useful to observe that the spaces $\cM_{\rm HM}$ arising from type II string compactifications (at string tree-level) are special in the sense that they admit a $2 n_H +1$-dimensional Heisenberg algebra of isometries. These shift symmetries reflect the gauge invariance of the ten-dimensional p-forms and remain unbroken by perturbative $g_s$-corrections. This implies in particular that there are $n_H+1$ commuting shift symmetries.\footnote{These $n_H+1$ shift symmetries are still preserved once the non-perturbative corrections from D(-1)- and D1-instantons in type IIB or their mirror configurations, A-type membrane instantons in type IIA are included.} These isometries can be made manifest by dualizing the corresponding hypermultiplet scalars into tensor fields, so that one deals with $3 n_H-1$ real scalars and $n_H+1$ rank two antisymmetric tensors. 

In type IIB compactifications ($n_H = h^{(1,1)}(Y) + 1$ with $h^{(p,q)}(Y)$ denoting the Hodge numbers of $Y$), the $h^{(1,1)}(Y) + 2$ tensor fields arise naturally as the space-time parts of the NS and RR two-form and the expansion of the RR four-form with respect to the $h^{(1,1)}$ harmonic two-forms of $Y$. The tensor multiplet spectrum is completed by 
 $3h^{(1,1)}(Y) + 2$ scalars comprising the ten-dimensional axion-dilaton $\tau = \tau_1 + \I \tau_2 = C_0 + \I \e^{-\phi}$ and $3 h^{(1,1)}$ scalars coming from the period integrals
of the K\"ahler form $J$ and the NS and RR 2-forms with respect to a basis of two-cycles $\gamma^a_{(2)} \in H^2(Y, \Zom)$
\be\label{IIBscalars}
z^a = b^a + \I t^a = \int_{\gamma^a_{(2)}} B_{\rm NS} + \I J \, , \qquad 
c^a = \int_{\gamma^a_{(2)}} C_2 \, , \qquad a = 2, \ldots , h^{(1,1)}(Y)+1 \, ,
\ee

Utilizing the tensor multiplet picture, $\cM_{\rm HM}$ can elegantly be described by combining projective superspace and superconformal calculus techniques. The underlying idea is that $\cM_{\rm HM}$ with $n_H+1$ commuting isometries has an  equivalent  formulation in terms of rigidly superconformal tensor multiplet (TM) Lagrangians \cite{Hitchin:1986ea}  
\be\label{cont}
\cL = \int d^2 \theta d^2 \bar{\theta} \oint_{\cC} \, \frac{{\rm d} \zeta}{2 \pi \I \zeta} \, H(\eta^I) \, , \qquad I = 0, \ldots, h^{(1,1)}(Y)+1 \, .
\ee
Here $\eta^I = v^I/\zeta + x^I - \zeta \vb^I$ denotes a $\cN = 2$ tensor superfield, $H(\eta^I)$ is a weakly homogeneous function of degree one (i.e., logarithmic terms are allowed) with no explicit  $\zeta$-dependence, $\cC$ is an arbitrary contour in the complex $\zeta$-plane, and $\int d^2 \theta d^2 \bar{\theta}$ is the superspace measure. Coupling the Lagrangian \eqref{cont} to conformal supergravity (Weyl multiplet) and carrying out the superconformal quotient determines $\cM_{\rm HM}$ \cite{dWRV,deWit:2006gn}.

When taken as a function of the scalar fields, $\cL(v^I, \vb^I, x^I)$ gives rise to the tensor potential \cite{deWit:2006gn}
\be\label{TP}
\chi = - \cL + x^I \frac{\partial \cL}{\partial x^I} \, ,
\ee
which also fixes $\cM_{\rm HM}$. In fact carrying out the Legendre transform by setting $w_I + \wb_I = \frac{\partial \cL}{\partial x^I}$ and taking $x^I = x^I(v, \vb, w + \wb)$,  $\chi$ becomes the K\"ahler potential on the hyperk\"ahler cone (or Swann bundle) over $\cM_{\rm HM}$ \cite{dWRV}. Describing $\cM_{\rm HM}$ through the functions $\cL, \chi$ is in close analogy to the vector multiplet sector whose couplings are completely encoded in $F, \cK$. 

In the superconformal formulation, the physical fields \eqref{IIBscalars} are given by (SU(2) and dilatation invariant) functions of the tensor multiplet scalars \cite{Neitzke:2007ke}
\begin{equation}\label{vocabIIB}
\tau = \frac{1}{(r^{0})^2} \left(\vr{0} \cdot \vr{1} + \I \, |
\vr{0} \times \vr{1} | \right)\ ,\qquad  z^a =
\frac{\eta_+^a}{\eta_+^1}\ ,\qquad c^a = \frac{(\vr{0} \times
\vr{1}) \cdot (\vr{1} \times \vr{a})}{|\vr{0} \times \vr{1}|^2} \, ,
\end{equation}
where $\vr{I} = \left[2 \, v^I, \, 2 \, \vb^I, \, x^I \right]$ with $\vr{I}\! \cdot \vr{J} = 2 v^I \vb^J + 2 v^J \vb^I + x^I x^J$
and $\eta^\Lambda_+=\eta^\Lambda(\zeta_+)$, $\Lambda = \{1,a\}$, where $\zeta_+ = \tfrac{1}{2 \vb^0} (x^0 - |\vr{0}|)$ is one of the roots of $\zeta\eta^0(\zeta)$. Working with the tensor potential has the virtue that (up to a conformal factor $r^0 = |\vr{0}|$) $\chi$ can  be completely expressed in terms of the physical fields \eqref{vocabIIB}, which considerably simplifies the physical interpretation of the quantum corrections. Moreover, the close relation between $\chi$ and the K\"ahler potential on the hyperk\"ahler cone over $\cM_{\rm HM}$ implies that (discrete) symmetries of $\cM_{\rm HM}$ as, e.g., the SL(2, $\Zom)$ invariance of the type IIB string, are reflected by the invariance of $\chi$.

Let us close this section by remarking that superconformal invariance of the TM Lagrangian implies that $\chi$ is subject to a set of partial differential equations. These are automatically solved if $\chi$ is obtained via the contour formulation \eqref{cont}, but can also be used to verify supersymmetry without resorting to $\cL$. For one HM one finds a single constraint on $\chi$ which is equivalent to the condition for four-dimensional quaternion-K\"ahler spaces with two commuting isometries found by Calderbank and Pedersen \cite{CP}. The three partial differential equations and metric arising in the two HM case have been worked out in \cite{Tensor:proc}. These supersymmetry constraints will probably be an important ingredient in demonstrating the uniqueness of the modular completion below.

\section{Instanton corrections to the hypermultiplet moduli space}
In order to determine the $g_s$-corrections to $\cM_{\rm HM}$, we follow the strategy outlined in Figure \ref{eins}. 
Building on the off-shell description of the c-map \cite{RVV} 
the perturbatively corrected hypermultiplet moduli space has been determined in \cite{RSV}
\be\label{L1loop}
\cL(v, \vb, x) = {\rm Im} \oint_{\cC_0} \, \frac{{\rm d} \zeta}{2 \pi \I \zeta} \,  \frac{F(\eta^\Lambda)}{\eta^0} \pm 
\frac{\I \, 2 \zeta(2) \, \chi_E }{(2 \pi)^3} \oint_{\cC_1}   \, \frac{{\rm d} \zeta}{2 \pi \I \zeta}  \, \eta^0 \, \ln(\eta^0) \, .
\ee
Here $F(\cdot)$ is the holomorphic prepotential of the T-dual type IIA vector multiplet geometry, $\eta^0$ is an additional TM acting as a conformal compensator and the contours $\cC_0$ and $\cC_1$ enclose the root $\zeta_+$ of $\zeta \eta^0$ and the branch cut between $\zeta = 0$ and $\zeta_+$ respectively \cite{Alexandrov:2007ec}. The first term encodes the classical c-map including the tree-level $\alpha^\prime$ corrections to $\cM_{\rm HM}$. The second term is the universal one-loop correction with the upper (lower) sign corresponding to type IIB (type IIA) strings and $\chi_E$ being the Euler number of the CY.

Substituting \eqref{L1loop} into \eqref{TP} yields the tensor potential underlying the perturbatively corrected $\cM_{\rm HM}$. For type IIB the corresponding expression naturally splits into a classical part $\chi_{\rm cl}$, a piece containing the perturbative $\alpha^\prime$ and $g_s$ corrections $\chi_{\rm pt}$, and $\chi_{\rm ws}$ capturing the world-sheet instanton contributions 
\be\label{chi1loop}
\begin{split}
\chi_{\rm cl} = & \, 4 \, r^0 \, \tau_2^2 \, \frac{1}{3!} \, \kappa_{abc} \, t^a \, t^b \, t^c \, , \qquad
\chi_{\rm pt} =  \frac{1}{(2 \pi)^3} \, r^0 \, \chi_E \, \left[ \zeta(3) \tau_2^{2} + 2 \zeta(2) \right] \, , \\
\chi_{\rm ws} = & - \frac{r^0 \tau_2^{2}}{(2 \pi)^3}
\sum_{ k_a  } n_{k_a} \left[ {\rm Li}_3(\e^{2 \pi \I k_a z^a}) + 2 \pi k_a t^a {\rm Li}_2(\e^{2 \pi \I k_a z^a}) + c.c.\right] \, .
\end{split}
\ee
Here $\kappa_{abc}$ are triple intersection numbers and $n_{k_a}$ the Gopakumar-Vafa invariants of $Y$. The one-loop correction  is suppressed by $\tau_2^{-2} = g_s^2$ compared to the three-level terms and gives rise to the second term in $\chi_{\rm pt}$.

In order to determine the D(-1)- and D1-brane instanton corrections we make use of the non-perturbative SL(2, $\Zom)$ invariance of the type IIB string. The modular transformation properties of the four-dimensional scalars are inherited from the ten-dimensional fields through the dimensional reduction \cite{BGHL}
\be\label{SL2Z}
\tau \mapsto \frac{a \tau +b}{c \tau + d} \, , \quad t^a \mapsto t^a |c\tau+d| \, , \quad b^a \mapsto d \,b^a + c \,c^a \, , \quad c^a \mapsto b \, b^a + a \, c^a \,.
\ee
The modular transformation of the conformal compensator $r^0 \mapsto r^0 |c\tau+d|$ is determined by lifting the action of SL(2, $\Zom)$ to superspace \cite{Berkovits:1995cb}.

Applying the transformations \eqref{SL2Z} to \eqref{chi1loop} one finds that $\chi_{\rm cl}$ is modular invariant while the $\alpha^\prime$ and $g_s$ corrections break the SL(2, $\Zom$) symmetry. Restoring SL(2, $\Zom)$ invariance by a modular completion of $\chi_{\rm pt}$ and $\chi_{\rm ws}$ gives \cite{Robles-Llana:2006is}
\be\label{IIBTP}
\begin{split}
\chi^{\rm IIB}_{(-1)} & = \frac{r^0 \tau_2^{1/2}}{2 (2 \pi)^3}\,
\chi_E \, \sideset{}{'}\sum_{m,n}\, \frac{\tau_2^{3/2}}{|m\tau +
n|^3}\ , \\
\chi^{\rm IIB}_{(1)} & = - \frac{r^0 \tau_2^{1/2}}{(2 \pi)^3}\,
\sum_{ k_a } n_{k_a} \sideset{}{'} \sum_{m,n} \frac{\tau_2^{3/2}}
{|m\tau + n|^3}\, \big( 1 + 2 \pi |m\tau + n|\, k_a t^a \big) \,
\e^{-S_{m,n}}\, .
\end{split}
\ee
Besides the $\alpha^\prime$ and perturbative $g_s$ corrections this result also includes the instanton corrections from  D(-1)- and D1-branes, respectively. In particular one notes the appearance of the $(m,n)$-string instanton action  
\be\label{Smn}
  S_{m,n} = 2\pi k_a \big( |m\tau + n|\, t^a - \I m\, c^a - \I n\, b^a
  \big)\ ,
\ee
where $m$ and $n$ are the units of D1 and fundamental string charge.

By the SYZ construction of mirror symmetry \cite{SYZ}, the D-brane instanton corrections in \eqref{IIBTP} are mirror to Euclidean D2-branes, wrapping so-called A-cycles of the mirror CY. The corresponding mirror map in the hypermultiplet sector is given by \cite{BGHL,membrane}
\begin{equation}\label{MM}
\phi_{{\rm IIA}}=\phi_{{\rm IIB}}\ ,\quad A^1 = \tau_1\ ,\quad A^a =
-(c^a - \tau_1 b^a)\ ,\quad   z^a_{{\rm IIA}} = z^a_{{\rm IIB}}\ .
\end{equation}
Here the $A^\Lambda$ arise as the periods of the ten-dimensional RR three-form over (electric) A-cycles, while $z^a_{\rm IIA}$ are the complex structure moduli of the mirror manifold $X$.\footnote{Note that these relations require a special choice of symplectic basis in $H^3(X, \Zom)$, which induces a distinction between A- and B-cycles based on their properties under mirror symmetry, see \cite{SYZ} for details.} Implementing the mirror map requires extracting the contributions from D(-1)- and D1-branes from \eqref{IIBTP} following Table \ref{t.1}.
\begin{table}[t]
\begin{center}
\begin{tabular}{ll|ll}
IIB HM & SL(2, $\Zom)$-invariant & IIA HM & composed from IIB terms \\ \hline
$\chi_{\rm cl}$ & =  $\chi_{\rm cl}$
& $\chi_{\rm tree}$ & = $ \chi_{\rm cl} + \chi_{\rm ws-pert} + \chi_{\rm ws-inst}$ \\
$\chi_{(-1)}$ & = 
$\chi_{\rm ws-pert} + \chi_{\rm loop} + \chi_{D(-1)}$ 
& $\chi_{\rm loop} $ & = $\chi_{\rm loop} $\\
$\chi_{(1)}$ & = $ \chi_{\rm ws-inst} + \chi_{\rm D1}$ & 
$\chi_{\rm A-D2}$ & = $ \chi_{\rm D(-1)} + \chi_{\rm D1}$
\end{tabular}
\end{center}
\parbox[c]{\textwidth}{\caption{\label{t.1}{Rearrangement of the type IIB tensor potential under mirror symmetry.}}}
\end{table}
The resulting correction term due to D2-brane instantons was then given in \cite{membrane}
\be\label{D2instres}
\chi^{\rm IIA}_{{\rm A-D2}} = - \frac{r^0 \tau_2}
{2\pi^2}\, \sum_{k_\Lambda} n_{k_\Lambda} \sum_{m \not = 0}
\frac{|k_\Lambda z^\Lambda|}{|m|} \,  K_1\big( 2 \pi \tau_2 \, |m
\, k_\Lambda z^\Lambda| \big) \, \e^{-2 \pi \I m k_\Lambda
A^\Lambda}\ .
\ee
Here, 
$k_{\Lambda} = \big( n, k_a\big)$, $z^\Lambda = \big( 1 , z^a \big)$, $A^\Lambda = \big( A^1 , A^a \big)$, and the sum over $k_a$ now includes the zero-vector $k_a=0$, but
$k_\Lambda =0$ is excluded. The type IIA instanton numbers are
\be\label{electvec} n_{(n,k_a=0)} = \tfrac{1}{2}\chi_E(X)\ ,\qquad n_{(n,k_a)}
=n_{k_a}(Y) \, \quad \text{as in type IIB} \, . \ee
Note that here $n_{k_a}(Y)$ are the Gopakumar-Vafa invariants of the \emph{mirror} CY. This indicates a deep connection between the properties of holomorphic two-cycles of $Y$ (counted by $n_{k_a}(Y)$) and special Lagrangian three-cycles of the mirror manifold $X$.

One particular limit of eq.\ \eqref{D2instres} is the so-called conifold limit considered by Ooguri and Vafa \cite{OV}. In this limit one particular complex structure modulus and the string coupling constant are send to zero, $z^\star \rightarrow 0$, $g_s = \tau_2^{-1} \rightarrow 0$, while keeping the product $\tau_2 z^\star = \lambda$ and all other complex structure moduli $z^\Lambda, \Lambda \not = \star$ finite. In this limit gravity decouples and \eqref{D2instres} gives rise to a four-dimensional hyperk\"ahler metric. It was then shown in \cite{SV} that the resulting hyperk\"ahler metric is precisely the one found in \cite{OV} by evoking symmetry and regularity arguments.

The instanton corrections discussed above, however, do not determine the full quantum corrected LEEA, as the corrections from D2-brane instantons wrapping B-cycles in type IIA (or their mirror configurations from the D3- and D5-brane instantons in type IIB) and the contributions from NS5-branes have not been included. These instanton corrections are beyond the scope of the tensor multiplet formalism of Section \ref{two}, as they will break (some of) the isometries made manifest by the tensors. Nevertheless, the superconformal formulation based on the K\"ahler potential on a generic hyperk\"ahler cone \cite{dWRV} is still applicable and expected to play a key role in completing the implementation of the duality chain in Figure \ref{eins}.

\subsection*{Acknowledgements}
Its a pleasure to thank B.\ de Wit, D.\ Robles-Llana, M.\ Ro\v{c}ek, U.\ Theis and S.\ Vandoren for fruitful collaboration. Fruthermore, I would like to thank the organizers of the European Physical Society HEP 2007 conference and the Fifth Simons Workshop in Mathematics and Physics for their hospitality and S.\ Alexandrov, T.\ Grimm, S.\ Sethi, A.\ Klemm and C.\ Vafa for interesting discussions. This work is partially supported by the European Commission Marie Curie Fellowship no.\ MEIF-CT-2005-023966 and the grant ``Structure of vacuum, topological strings and black holes'' from the French {\it Agence Nationale de la Recherche}. 


\end{document}